%% file: loss_channels_arXiv.tex
\documentclass[aps,reprint,groupedaddress,amssymb,amsfont,onecolumn,notitlepage]{revtex4-1}

\usepackage{graphicx}
\graphicspath{{figures/}} 

\usepackage[T1]{fontenc}
\usepackage{lmodern}
\usepackage[utf8]{inputenc}
\usepackage[english]{babel}
\usepackage{todonotes}
\usepackage{textcomp} 
\usepackage{soul,xcolor}
\setstcolor{red}

\newcommand{\gnot}{g$_{\text{0}}$~}

\begin{document}
	
	\title{Efficient anchor loss suppression in coupled near-field optomechanical resonators}
	
	\author{Luiz, G. O.}
	\affiliation{Gleb Wataghin Physics Institute, University of Campinas, 13083-859 Campinas, SP, Brazil}
	
	\author{Santos, F. G. S.}
	\affiliation{Gleb Wataghin Physics Institute, University of Campinas, 13083-859 Campinas, SP, Brazil}
	
	\author{Benevides, R. S.}
	\affiliation{Gleb Wataghin Physics Institute, University of Campinas, 13083-859 Campinas, SP, Brazil}
	
	\author{Espinel, Y. A. V.}
	\affiliation{Gleb Wataghin Physics Institute, University of Campinas, 13083-859 Campinas, SP, Brazil}
	
	\author{Alegre, T. P. Mayer}
	\affiliation{Gleb Wataghin Physics Institute, University of Campinas, 13083-859 Campinas, SP, Brazil}
	
	\author{Wiederhecker, G. S.}
	\email[]{gustavo@ifi.unicamp.br}
	\affiliation{Gleb Wataghin Physics Institute, University of Campinas, 13083-859 Campinas, SP, Brazil}
	
	\date{\today}
	
	\begin{abstract}
		Elastic dissipation through radiation towards the substrate is a major loss channel in micro- and nanomechanical resonators. Engineering the coupling of these resonators with optical cavities further complicates and constrains the design of low-loss optomechanical devices. In this work we rely on the coherent cancellation of mechanical radiation to demonstrate material and surface absorption limited silicon near-field optomechanical resonators oscillating at tens of MHz. The effectiveness of our dissipation suppression scheme is investigated at room and cryogenic temperatures. While at room temperature we can reach a maximum quality factor of 7.61k ($fQ$-product of the order of $10^{11}$~Hz), at 22~K the quality factor increases to 37k, resulting in a $fQ$-product of $2\times10^{12}$~Hz.
	\end{abstract}
	
	
	\maketitle
	
	\section{Introduction}
	
	The interaction of optical and mechanical fields in microscale devices enables the manipulation and control of mechanical modes vibrating at radio-frequencies. Some remarkable examples resulting from such an optomechanical interaction include the preparation and measurement of harmonic oscillators' quantum ground states~\cite{chan_laser_2011,Ying2014,Ludwig2012}, optically induced synchronization between mechanical oscillators~\cite{zhang_synchronization_2012-1}, phase noise suppression~\cite{Zhang2015b} and highly sensitive sensors~\cite{Liu2013,Cervantes2013a,Reinhardt2016,Norte2016}. A major limitation in these microdevices is mechanical energy loss that leads to reduced sensitivity~\cite{huang_optomechanical_2013}, lower coherence~\cite{Aspelmeyer2014}, and increased power consumption~\cite{kippenberg_analysis_2005}; it also remains among the most challenging issues in the design and fabrication processes of micromechanical devices.
	
	While mechanical dissipation is ultimately limited by material absorption, it may also depend on the viscosity of the environment gas, anchor losses and surface scattering or absorption. Although the gas damping can be suppressed in vacuum, the necessary anchoring of micromechanical devices leads to radiation of mechanical waves towards the substrate, which is often the dominant dissipation channel\cite{Cole2011}. Fortunately, anchor dissipation can be reduced through phononic shielding~\cite{alegre_quasi-two-dimensional_2011,Hsu2011,Tsaturyan2016}, tensile-stressed materials~\cite{Verbridge2006, Norte2016} , mesa isolation~\cite{Pandey2009}, and destructive interference of elastic waves~\cite{Zotov2014,Zhang2014a,wilson-rae_high-q_2010}. Phononic shields are very effective for high frequency mechanical modes\cite{Santos2016a,Benevides2017}, but their very large footprint at low frequencies\cite{Tsaturyan2016} is not optimum for photonic integrated circuits. The mesa approach is efficient only for out-of-plane mechanical modes and its deep etch of the substrate may also be incompatible with photonic integration. Therefore, although very high $Q$ mechanical resonators have been reported, they  can not always be easily  integrated with optical cavities using large-scale photonic integration technology. On the other hand, exploring the destructive interference of mechanical waves is a simple method to obtain both low~\cite{Zhang2015a} and high~\cite{Sun2012} frequency high-$Q$ mechanical resonators, without impacting the device's design or footprint. Yet, the constraints of simultaneously supporting optical and mechanical modes still challenge optomechanical device's design. 
	
	Near-field optomechanical (NFO) devices~\cite{Anetsberger2009,Srinivasan2011,kim_nanoscale_2012,Doolin2014a} can overcome these challenges as their mechanical and optical resonators are separated structures, which interact through the evanescent optical field. Here we demonstrate a silicon NFO device, fabricated in a commercial foundry, that can reach surface-limited $Q$-factors by efficient suppression of anchor losses through elastic wave interference.
	
	\section{Device presentation and anchor loss suppression}
	
\begin{figure}[htb]
		\center
		\includegraphics[scale=1]{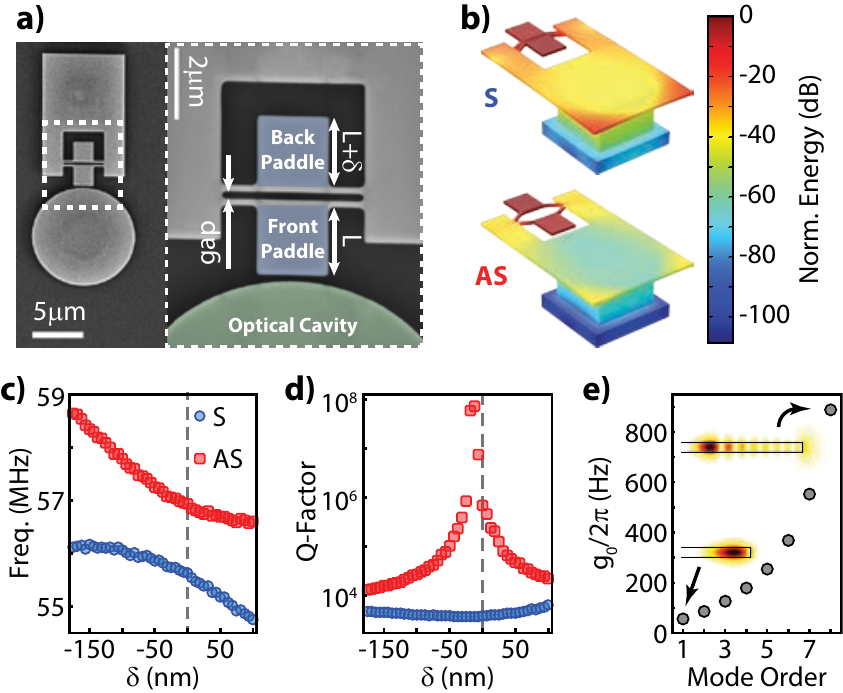}
		\caption{\label{fig:1}\textbf{a)} Scanning electron microscopy of the device. \textbf{b)} Finite element method (FEM) simulation of the normalized energy distribution of the mechanical modes of interest for the device with $\delta=-15$~nm in (\textbf{c},\textbf{d}); displacement is exaggerated and colors indicate normalized energy distribution. \textbf{c,d)} Mechanical frequency and $Q$-factor dependence on paddles balance ($\delta$) from FEM simulations. \textbf{e)} Perturbation theory estimate for the optomechanical coupling for different radial order optical TE modes with $\lambda=1520$~nm (see appendix B for more information); insets: $|E_r|$ distribution profiles of the first (bottom) and $8^{th}$ (top) radial order TE modes; darker colors correspond to higher intensities. In (\textbf{b}-\textbf{d}) only radiation to the substrate is considered.}
	\end{figure}

	Our NFO device is based on a mechanical resonator composed of coupled paddles that interact with a nearby silicon microdisk optical cavity (Fig.~\ref{fig:1}a). This design allows us to perfectly balance the mechanical waves radiated to the pedestal by each paddle, without changes to the optical cavity design. Furthermore, it sets an interesting platform to study optomechanical arrays, as it allows coupling several mechanical resonators to a single optical cavity~\cite{xuereb_strong_2012,Xuereb2014,holmes_multi-stability_2011}.
	
	The mechanical resonator is composed by two square paddles (2~\textmu m$~\times$~2~\textmu m), attached on both sides to suspended beams through 4 nanostrings (200~nm wide) separated by a 200~nm gap. The device is defined on a 220~nm silicon-on-insulator (SOI). Because the paddles' motion couple through the supporting beams, symmetric (S) and anti-symmetric (AS) combinations of individual paddle modes are formed, such as the in-plane modes shown in the finite element method (FEM) numerical simulation of Fig.~\ref{fig:1}b. In order to reach a perfect balance between mechanical waves radiating to the supporting beams, the length of the back paddle ($L$) is offset from the front one by a small length $\delta$. FEM calculations of the mechanical modes show that the resonant frequencies of the coupled mechanical modes display the avoided crossing behavior when $\delta$ is varied (Fig.~\ref{fig:1}c), a signature of coupling of the paddles' motion. The calculations also show that the AS mode has consistently higher anchor loss limited $Q$'s when compared to the S mode (Fig.~\ref{fig:1}d). This difference appears because the S mode induces a larger displacement on the supporting beams, due to the two paddles' in-phase motion, coupling energy from the paddles to the pedestal and leading to a higher loss rate. On the other hand, the AS mode, due to the anti-phase paddle motion, drastically reduces the displacement at the anchor points, minimizing dissipation to the substrate when the two radiated mechanical waves are balanced. Moreover, we observe a rapid increase of this effect when the paddles are more symmetric ($\delta\approx 0$), indicating that it is possible to eliminate anchor losses in such NFO device simply by balancing the paddles. Note however that, due to geometry asymmetry of the single-sided pedestal supporting the beams, simulations show that the point of minimum dissipation of the AS mode happens with $\delta\approx -15$~nm.
	
	The devices were fabricated through the EpiXfab initiative at IMEC. They are defined on the 220~nm top silicon layer by deep UV (193 nm) photolithography and plasma etching at the foundry. The 2~\textmu m buried oxide is then partially removed using wet-etch (Buffered Oxide Etch) to define the pedestal and grant the device its mechanical degrees of freedom; this post-process step is performed in-house. Each die has a series of devices where the back paddle has its length varied, while the front paddle is fixed.
	
	\section{Measurement scheme and optical characterization}
	
	In order to readout the motion of the paddles, the front paddle is designed to be 200~nm away from a 5~\textmu m radius disk optical cavity (Fig.~\ref{fig:1}a) supporting whispering gallery modes~\cite{Anetsberger2009} (WGM). Optical readout is possible because the motion of the paddles modulates the frequencies of the WGM's through evanescent field perturbation. The figure of merit of this interaction is the optomechanical coupling rate~\cite{Aspelmeyer2014}, $g_0=(\partial\omega/\partial x) x_{\text{zpf}}$, which measures the amount of optical frequency shift caused by a displacement with a quantum-mechanical zero-point fluctuation amplitude ($x_{\text{zpf}}=\sqrt{\hbar/4 \pi m_{\text{eff}} f_m}$, where $\hbar$ is the Planck's constant, $m_{\text{eff}}$ the effective mass and $f_m$ the resonance frequency). Although there are usually two main contributions to the optical resonance shift, namely boundary motion~\cite{johnson_perturbation_2002} and photo-elastic effect~\cite{Chan2012}, we consider only the former because the mechanical modes we are interested induce negligible strain throughout the paddle volume. We estimated $g_0$ of the in-plane modes by employing perturbation theory~\cite{johnson_perturbation_2002} to calculate $\partial\omega/\partial x$ for the various radial orders of the optical transverse electric (TE) modes~\textemdash~  also calculated using FEM. The $g_0$ values -- for the perfectly balanced device -- ranges from tens to several hundred Hz, as shown in Fig.~\ref{fig:1}e (see appendix B for more information).
	
\begin{figure}[htb]
		\center
		\includegraphics[scale=1]{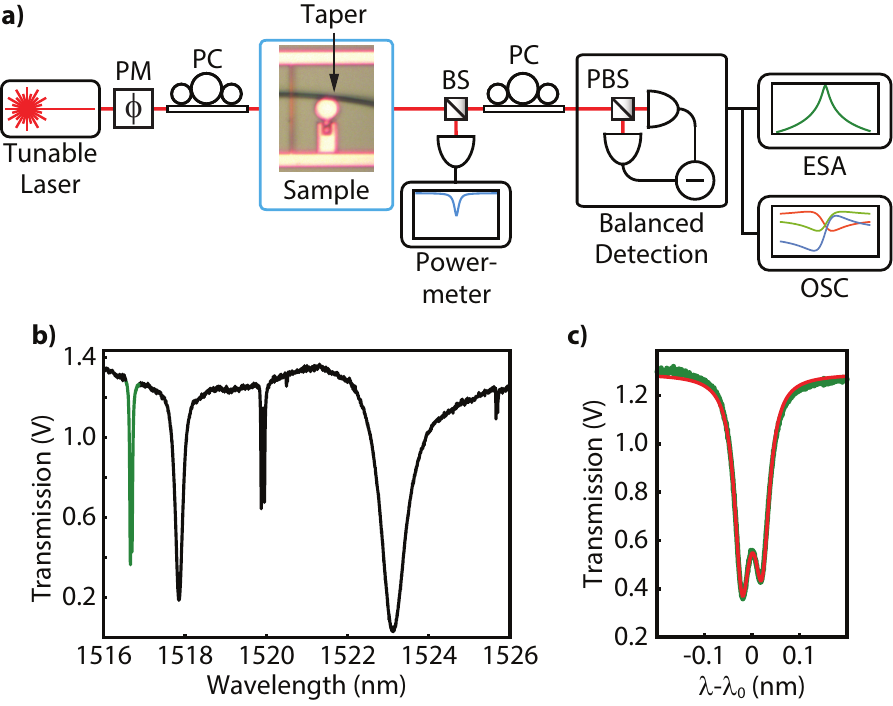}
		\caption{\label{fig:2}\textbf{a)} Schematics of the experimental setup. \textbf{b)} Optical spectrum of a typical device - marked in green is the resonance used to probe the mechanical modes. \textbf{c)} Measured resonance (green) and fitted counter-propagating coupled modes function (red) - $\lambda_0=1516.68$~nm and loaded quality factor $Q_{opt}=40$k. In (\textbf{a}): PM = phase-modulator; PC - polarization control; BS = beam splitter; PBS = polarizing beam splitter; ESA = electric spectrum analyzer; OSC - oscilloscope. In \textbf{c}: fitting (red) of the peak marked in \textbf{b} (green).}
	\end{figure}

	In order to probe the devices we couple light from an external cavity tunable laser into the disk resonator through a tapered fiber (Fig.~\ref{fig:2}a). The transmitted signal reveals the microdisk optical resonances (Fig.~\ref{fig:2}b), which usually exhibit frequency splitting due to counter-propagating optical mode coupling, induced by surface roughness and paddle-induced scattering\cite{Little1997a} (Fig.~\ref{fig:2}c). Nevertheless, these modes present reasonably high loaded optical quality factors of about $Q_{\text{opt}}=40$k, similar to other devices fabricated in the same foundry\cite{Santos2016a}. The optical mode used to probe the mechanical motion is chosen by monitoring the radio-frequency power spectrum strength at the resonant mechanical frequency, which depends on $g_0$, optical linewidth, and taper-cavity loading conditions\cite{Gorodetsky2010}. We identify which optical mode is excited by comparing the measured optical free spectral range (FSR) with FEM simulations over a broad range ($1460$~nm to $1610$~nm~\textemdash~see appendix A for more information), which indicates that we are coupling light to a higher radial order TE modes; indeed, good agreement between experimental and theoretical values for $g_0$ also supports this assumption (Fig.~\ref{fig:1}e). Measurement of $g_0$ and calibration of the PSD into a displacement noise spectrum density [m/Hz$^{1/2}$] are performed using a calibrated phase-modulation tone close to the mechanical resonances~\cite{Gorodetsky2010}. The mechanical modes are identified by directly comparing measured frequencies with those from FEM simulations, which agree within a $2\%$ margin. All measurements are performed with the optical cavity undercoupled to the taper.
	
	Using a homodyne detection scheme~\cite{Hansch1980} we are able to probe the samples with the laser tuned to the center of the optical resonance, thus avoiding any optomechanical backaction that could affect the mechanical quality factor. With the \gnot measured on the experiments ($\approx 2 \pi \times 450$~Hz) we estimate that even for a 10\% of optical linewidth deviation from the resonance we would obtain only a 0.3\% error in our mechanical Q's, assuming that all the light is coupled into the cavity (critical coupling with correct polarization). Because we control the polarization of the incident light such that there is very little light inside the cavity, the optomechanical feedback in case of laser deviation from resonance is even more negligible.
	
	\section{Results and discussion}
	
\begin{figure}[htb]
		\center
		\includegraphics[scale=1.0]{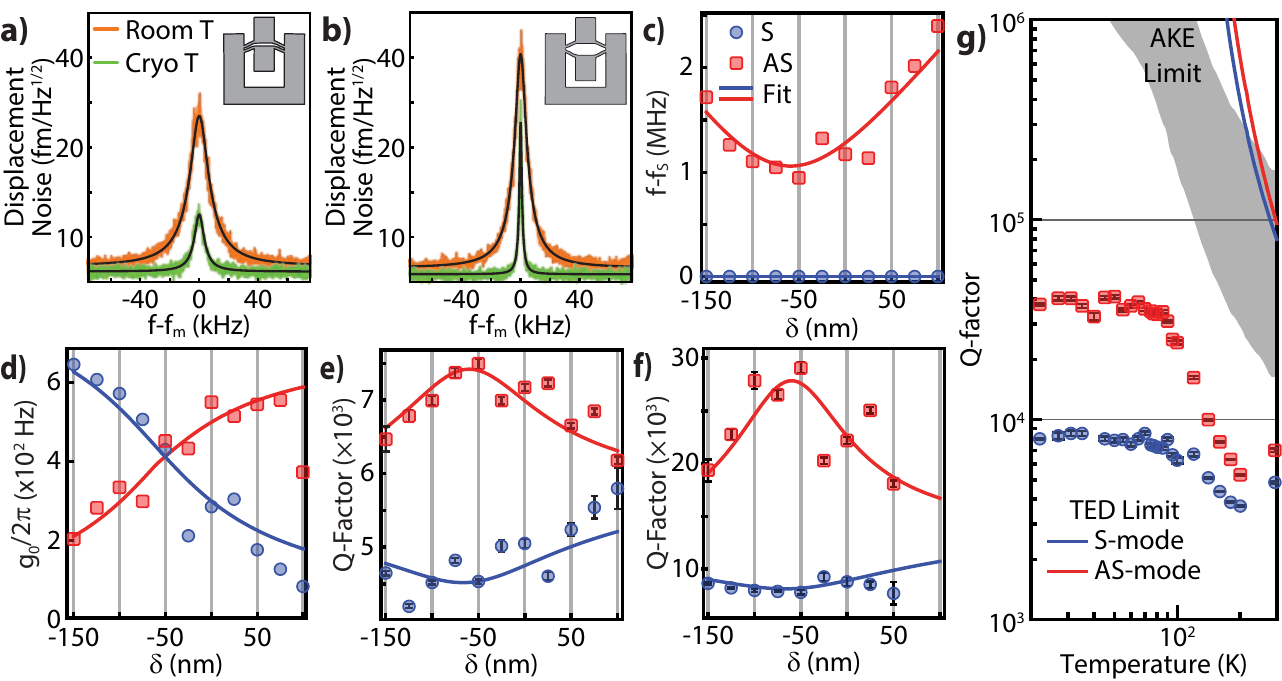}
		\caption{\label{fig:3} \textbf{a,b)} S (\textbf{a}) and AS (\textbf{b}) modes calibrated power spectrum density at room (orange) and cryogenic (green) temperatures; data in colors and fitted Lorentzians in black; $f_m\approx56$~MHz. \textbf{c-f)} Asymmetry dependence of frequency difference between S (blue) and AS (red) modes at room temperature (\textbf{c}), $g_0$ measured at room temperature (\textbf{d}), $Q$-factor at room (\textbf{e}) and cryogenic (\textbf{f}) temperatures; marks are data while lines are fitted curves of the coupled oscillators model. \textbf{g)} Temperature dependence of quality factor in log~$\times$~log scale; blue and red lines are the estimated S and AS modes TED limited $Q$'s; the gray area is bounded by the upper and lower AKE limits. In \textbf{a},\textbf{b},\textbf{g}): data of device with $\delta=-50$~nm.}
	\end{figure}

	The room temperature (RT) mechanical quality factors ($Q_m$) are obtained  from the measured power spectral density (PSD) of the two in-plane mechanical modes while the sample is in a vacuum chamber ($10^{-5}$~mbar). In Fig.~\ref{fig:3}(a,b) (orange curves) we show the measured PSD's and their fitted Lorentzians for the device with $\delta=-50$~nm, which has the highest AS mode quality factor, $Q_{m,\text{AS}}^{\text{RT}}=(7.61\pm0.07)$k, and $Q_{m,\text{S}}^{\text{RT}}=(4.53\pm0.04)$k for the S mode. The mechanical resonance frequencies of these coupled modes are around $f_{m}=56$~MHz with a frequency splitting of $\Delta f_{m}=940$~kHz, also in good agreement with the FEM simulations; this corresponds to a $fQ$-product of $4\times10^{11}$~Hz. A confirmation that the $\delta=-50$~nm device is indeed the one with best anchor loss suppression can also be drawn from the frequency and optomechanical couplings measured for devices with varying balance between paddles, as shown in Fig.~\ref{fig:3}(c,d). This device not only has the smallest frequency difference but it also has S/AS modes with almost identical measured optomechanical coupling rates ($g_0\approx 2 \pi \times 450$~Hz), which is expected due to its similar masses and frequencies of S/AS modes for the most balanced device.
	
	Comparison of the S and AS modes' quality factors, for devices with different $\delta$ (Fig.~\ref{fig:3}e), shows consistently higher quality factors for the AS modes, suggesting an important contribution from anchor loss. Nevertheless, its modest two-fold improvement compared to the S mode indicates that other loss mechanisms are also playing an important role in the overall dissipation.
	
	In order to assess the role played by the temperature dependent dissipation mechanisms, we insert the sample in a cold-finger cryostat to cool it down to 22~K. At these low temperatures (LT), we observe a high enhancement of 385\% for the AS mode quality factor, reaching $Q_{m,\text{AS}}^{\text{LT}}=(37.0\pm0.6)$k (Fig.~\ref{fig:3}b, green curve), while the S mode increases only by 80\%, up to $Q_{m,\text{S}}^{\text{LT}}=(8.2\pm 0.1)$k (Fig.~\ref{fig:3}a, green curve), for the same device with $\delta = -50$~nm. These $Q$-enhancements are accompanied by a small (0.7\%) increase in mechanical frequencies due to an expected material stiffening at LT, resulting in a $Qf$-product of $2\times10^{12}$~Hz.
	
	Such a high contrast between the S/AS modes' quality factors at LT clearly indicates the efficiency of the destructive interference scheme. Nonetheless, the calculated anchor loss limited $Q_m$ is still much higher than the measured values for the AS mode at LT. This suggests that either destructive interference between the mechanical waves is unbalanced, intrinsic material dissipation, such as thermoelastic damping (TED)~\cite{Lifshitz2000} or  Akhiezer effect (AKE)~\cite{Akhiezer1939,Woodruff1961a}, was reached or surface-related dissipation~\cite{Yasumura2000,Evoy2000} is playing an important role at 22~K. 
	
	The Q-limit imposed by destructive interference should not depend on temperature and its possible failure is assessed using a simple analytical model of three coupled mass-spring lumped oscillators~\cite{Zhang2014a}. The fitted model, which is detailed in the appendix C, predicts not only the frequency and damping dependence on the asymmetry $\delta$, but also takes into account an overall temperature-dependent intrinsic loss (e.g., material and surface losses). The solid lines in Fig.~\ref{fig:3}(c-f) represent the fitted model prediction, resulting in an fitted anchor loss, for both LT and RT data, that agrees within 1\%, while the intrinsic loss varies by a factor 4 (see appendix C for more details). Therefore our model indicates that, despite the similar fitting of LT and RT anchor loss, the observed sharpening in the LT data is reasonably explained by a reduction of the LT intrinsic damping. Also, the slowly varying AS mode $Q$-enhancement \textendash~towards the higher symmetry region ($\delta\approx-50$~nm) \textendash~suggests that the $Q$ is not being limited by variations in the device geometry. Indeed, numerical simulations confirm that small ($\pm100$~nm) variations on any of the device transverse dimensions would not quench the AS quality factors to the LT measured levels. Hence, we infer that the AS mode LT $Q$-factor is not being limited by failure of the destructive interference scheme. 
	
	Intrinsic material dissipation, such as the thermoelastic damping (TED, caused by heat flow between mechanically heated/cooled regions) and the Akhiezer effect (AKE, caused by strain-induced perturbation of thermal phonon equilibrium), have a distinct temperature-dependence~\cite{Zener1937,Lifshitz2000,Akhiezer1939,Woodruff1961a}. Therefore, we investigate their role by increasing the cryostat base temperature from 22~K up to 200~K, while monitoring the mechanical quality factors (Fig.~\ref{fig:3}g). The measured $Q$ temperature dependence reveals a capped behavior below $\sim100$~K and a power-law reduction up to $\sim200$~K; above this temperature there is an apparent increase in both S/AS quality factors.
	
	The large mismatch between the period of the mechanical oscillation ($~20$ ns) and the thermal lifetimes ($~\mu$s) in these devices results in very small TED dissipation. From FEM calculations, shown in Fig.~\ref{fig:3}g (blue and red lines), we expected room-temperature $Q$ limit to be around $10^5$, while the expected low temperature TED-limited $Q$'s are well above the $10^8$ level, without ever decreasing below the RT limit (see appendix D for more details). Hence TED cannot explain neither the temperature dependency nor the LT and RT $Q$ limits.
	
	The AKE, while irrelevant at low-temperatures, could be playing an important role at room-temperatures. We verify it by considering the model by Woodruff and Ehrenreich\cite{Woodruff1961a} with temperature dependent material properties \cite{Desai1986,Lambade1995,Philip1983} (see appendix E for the used values). Among the parameters involved in this model, the Gr\"uneisen parameter ($\gamma$) has the largest range of reported values~\cite{Ghaffari2013}; room-temperature values range from $\gamma^{RT} = 0.17$ up to $\gamma^{RT} = 1.5$, resulting in an upper and a lower limit for the AKE, respectively. The gray area in Fig.~\ref{fig:3}g is bounded by these upper and lower limits, and the lower-limit is roughly within a factor 2 above the RT measured values. Despite this mismatch and the large uncertainty in the Gr\"uneisen parameter, $Q$-factor temperature dependence follows the AKE power-law, which suggests the Akhiezer effect also plays an important role in the overall damping close to room temperature.
	
\begin{figure}[htb]
		\center
		\includegraphics[scale=1]{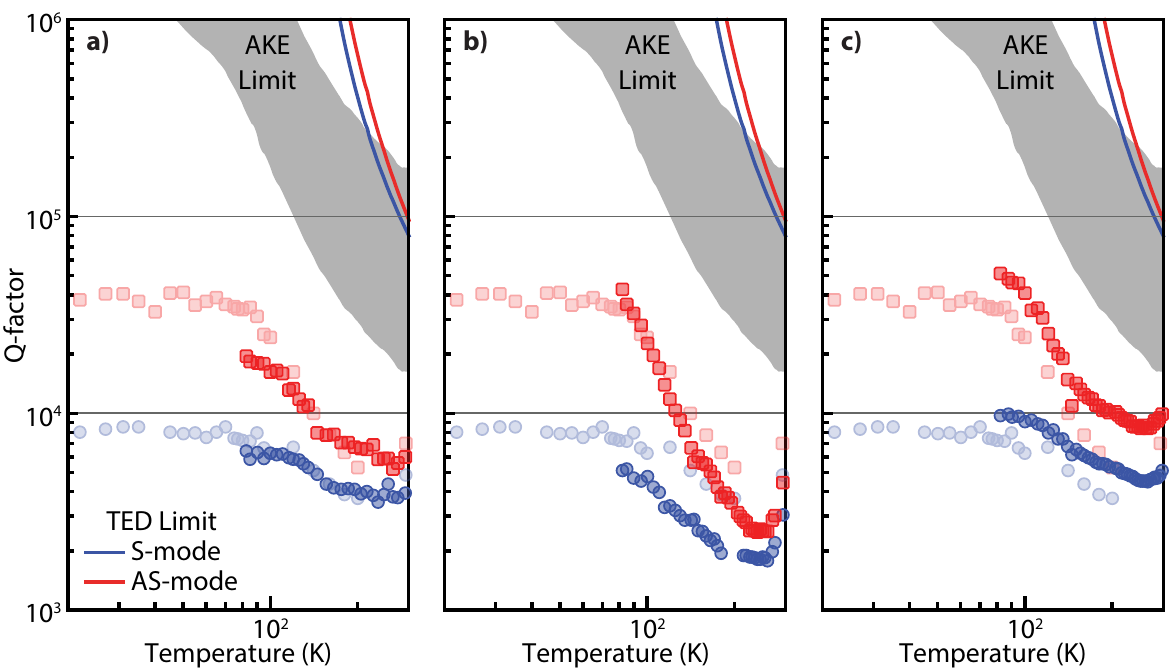}
		\caption{\label{fig:4}\textbf{Q-factor of S (blue) and AS (red) modes for different situations.} \textbf{a)} After a few months stored in a box with N$_{2}$ rich atmosphere; \textbf{b)} After cleaning with piranha and HF dip; \textbf{c)} After in-situ laser annealing~\textemdash~device with $\delta=-75$~nm. We keep the first cool down measure in every figure for comparison (lighter colors). Gray area marks the Akhiezer limits and the blue and red lines the TED limits.}
	\end{figure}

	Surface-related dissipation, scattering and absorption, are known to be significant in thin silicon devices~\cite{Yasumura2000,Evoy2000}, and is found here to play a role across the whole measured temperature range. The first hint of these surface-effects at LT arises from the data shown in Fig.~\ref{fig:3}f, which was actually taken during a second cool-down of the sample. Its maximum $Q$ is 35\% smaller than that shown in Fig.~\ref{fig:3}g (first cool-down), suggesting that some sort of surface modification occurred between measurements. We verified this assumption by measuring the impact of surface treatment of our samples \textemdash~chemical cleaning\cite{Borselli2006} (Fig.~\ref{fig:4}b) and local laser annealing\cite{Aubin2003} (Fig.~\ref{fig:4}c)~\textemdash~on the $Q_m$ temperature-dependence. Up to 45\%  improvement over the highest $Q_m$ shown in Fig.~\ref{fig:3}f was observed for the better balanced device ($\delta=-50$~nm) after cleaning. Other tested devices have also shown similar improvement on the quality-factor, as observed on the measurements of the device with $\delta = -75$~nm in Fig.~\ref{fig:4}c. Although further investigation would be required to further reduce surface losses, the observed $Q$-enhancement after surface treatment confirm that the LT $Q$-limit is unlikely due to failure of the destructive interference effect. The second hint suggesting surface losses at higher temperatures arises from the consistent $Q$-minimum around 220~K, which is coarsely observed in Fig.~\ref{fig:3}g, but consistent in all liquid-nitrogen cool-downs (Fig.~\ref{fig:4}). Similar minima have been observed in other structures\cite{Khiznichenko1967,Yasumura2000,Evoy2000,Li2009b,Tao2014,Tao2015} and, although typically associated with dislocation relaxation\cite{Khiznichenko1967} or surface related effects\cite{Yasumura2000,Evoy2000}, their actual origin still lack proper explanation and its detailed investigation lies beyond the scope of our paper.
	
	\section{Conclusion}
	
	In summary, we have demonstrated the use of destructive interference of elastic waves as an effective approach to obtain high-$Q$ near-field optomechanical resonators using a scalable foundry technology. Our data clearly shows that symmetric modes are highly susceptible to anchor loss, while the antisymmetric modes may be limited by other mechanisms, such as surface  and Akhiezer effect. To further increase these devices' performance, dedicated surface treatment steps should be carefully considered during the fabrication process~\cite{Khiznichenko1967,Tao2016,Tao2015,Aubin2003}, which  plays an important role across all temperature ranges. We expect these devices to serve as platforms for studying arrays of optically coupled mechanical resonators, as well as very sensitive force sensors.
	
	\section*{Appendix}
	\renewcommand\thesection{\Alph{section}}
	\setcounter{section}{0}
	\section{\label{sec:S1.1}Optical mode polarization and radial order}
	
\begin{figure}[hbt]
		\center
		\includegraphics[scale=1.0]{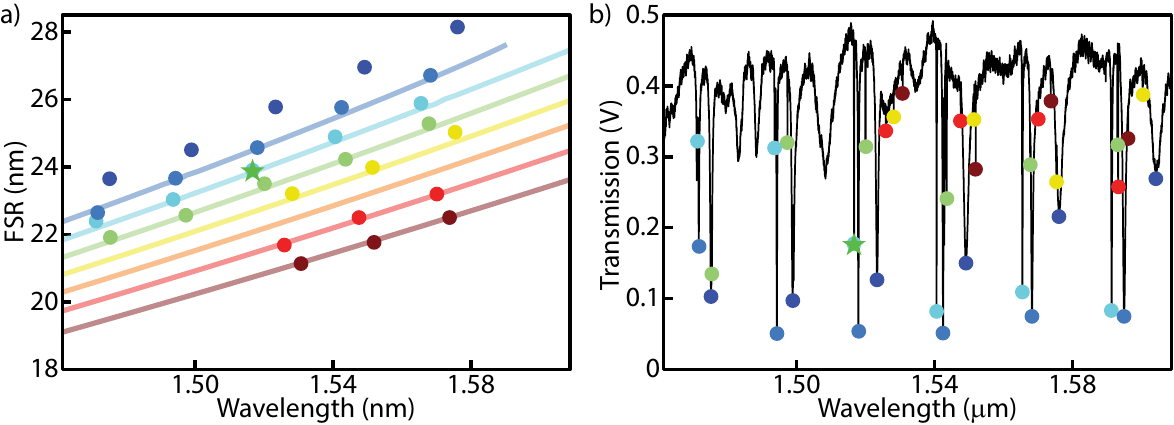}
		\caption{\label{fig:S1} \textbf{a)} Dispersion of experimental (colored circles) FEM calculated (solid lines) TE-like optical modes. \textbf{b)} Experimental optical spectra for the TE-like polarization. Circles on \textbf{a} and \textbf{b} are related by color. The green star marks the optical mode used to probe the mechanical resonator. In \textbf{a} the FEM calculated modes increase radial order from bottom to top (1$^{st}$ to 7$^{th}$).}
	\end{figure}

	We determined the polarization of the optical modes (TE/TM) by comparing finite element method (FEM) numerical solutions of the disk's modes to the measured spectra. We use FEM to solve the modes' azimuthal numbers ($m$) for various wavelength values ($\lambda_0$) and calculate their separation (free-spectral-range or FSR). This gives us the dispersion curves shown as solid lines in Fig.~\ref{fig:S1}a, where the radial order increases from bottom up. Then we determine the experimental families of modes from the experimental transmission data, calculate their FSR and plot their dispersion (FSR vs. $\lambda_0$) over the numerical solution; these are shown as dots in Fig.~\ref{fig:S1}. We marked the optical mode used to probe the mechanical resonators with a green star.
	
	We calibrate the wavelength vector of the experimental data using a Mach-Zehnder interferometer and an acetylene cell. This results in very good agreement with the numerically calculated dispersion of a 4.9~\textmu m radius silicon disk, which is only 2\% smaller than the nominal disk, below any fabrication precision. We repeated the procedure with TM-like numerical and experimental data (not shown) which also agree very well.
	
	We note that the fact that there are families of modes with smaller FSR than the one used to probe the mechanical resonator (green star) is evidence that we in fact use a high radial order mode to perform the measurements.
	
	\section{\label{sec:S1.2}Dependence of \gnot on the optical mode radial order}
	
	The moving boundary optomechanical coupling rate (\gnot) can be estimated through perturbation theory with equation \ref{eq:S1}~\cite{johnson_perturbation_2002}:
	\begin{equation}
	\label{eq:S1}
	\frac{\text{g}_{\text{0}}}{x_{zpf}} = \frac{\omega_0}{2}\frac{\int_{S} |\textbf{U}_{n}| \left( \Delta \epsilon |\textbf{E}_{t}|^{2}+\Delta\epsilon^{-1} |\textbf{D}_{n}|^{2} \right) dA}{\int_{V}\epsilon_{0} n^{2} |\textbf{E}|^2 dV}
	\end{equation}
	\noindent
	where $x_{zpf} = \sqrt{\hbar/(2m_{eff}\Omega_m)}$ is the quantum zero-point fluctuation of a mechanical mode with effective mass $m_{eff}$ and frequency $\Omega_{m}$, $\omega_0$ is the optical unperturbed resonance frequency, $\textbf{U}_{n}$ is the normalized mechanical displacement perpendicular to the surface $S$, $\Delta \epsilon = \epsilon_{0}(n_{in}^{2}-n_{out}^{2})$ is the difference of refractive index inside and outside the material, $\Delta \epsilon^{-1} = \epsilon_{0}^{-1}(n_{in}^{-2}-n_{out}^{-2})$ is the difference of $n^{-2}$ inside and outside the material, $\textbf{E}_{t}$ is the field tangent to the surface, $\textbf{D}_{n}$ is the electric displacement normal to the surface and $\textbf{E}$ is the total electric field. 
	
	We recall that the optical effective mode volume can be defined as~\cite{Robinson2005a}:
	\begin{equation}
	\label{eq:S2}
	V_{eff} = \frac{\int_{V}\epsilon_{0} n^{2} |\textbf{E}|^2 dV}{\epsilon_{0} n_{max}^{2} |\textbf{E}|_{max}^{2}}
	\end{equation}
	\noindent
	where $|\textbf{E}|^{2}_{max}$ is the maximum field intensity and $n_{max}$ is the refractive index of the medium at the point where the field is maximum.
	
	We may then rewrite equation \ref{eq:S1} as:
	\begin{equation}
	\label{eq:S3}
	\frac{\text{g}_{\text{0}}}{x_{zpf}} = \frac{\omega_0 \rho}{2 \epsilon_{0} n^{2}_{max} V_{eff}}
	\end{equation}
	\noindent
	where we defined
	\begin{equation}
	\label{eq:S4}
	\rho = \int_{S} |\textbf{U}_{n}| \left( \Delta \epsilon |\tilde{\textbf{E}}_{t}|^{2}+\Delta\epsilon^{-1} |\tilde{\textbf{D}}_{n}|^{2} \right) dA
	\end{equation}
	\noindent
	and $|\tilde{\textbf{E}}_{t}|^{2} = |\textbf{E}_{t}|^{2} / |\textbf{E}|_{max}^{2}$ and $|\tilde{\textbf{D}}_{n}|^{2} = |\textbf{D}_{n}|^{2} / |\textbf{E}|_{max}^{2}$.
	
\begin{figure}[htb]
		\center
		\includegraphics[scale=1.0]{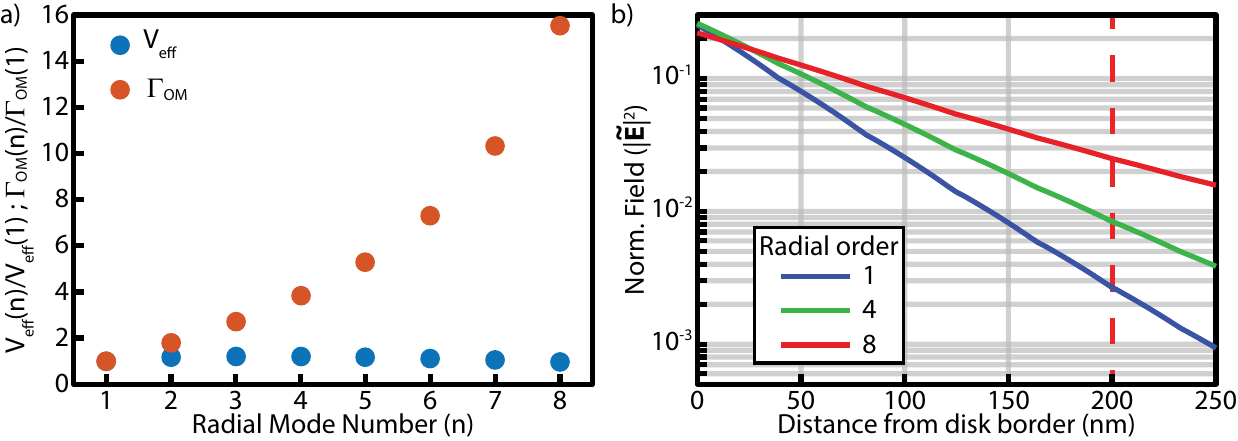}
		\caption{\label{fig:S2} \textbf{a)} Effetive mode volume (V$_{eff}$) and $\rho$, calculated at 200~nm away from the disk's border, for different radial order modes, normalized by the value for the first order mode. \textbf{b)} FEM calculated evanescent field profile, for three different radial order modes, outside of a 5~\textmu m radius and 220~nm thick Silicon disk optical cavity. The field in \textbf{b} is normalized by the maximum field value for each mode, which occurs inside of the cavity; the dashed vertical line indicates the position of the paddle closest to the disk.}
	\end{figure}

	In equation \ref{eq:S3} only $\rho$ and $V_{eff}$ depend on the field distribution, hence they must be the terms that determine the dependence of \gnot on the modes' order. Using FEM simulations, solving for the azimuthal number~\textemdash~same as solving for the effective index in a wave-guide~\textemdash~and maintaining the wavelength constant, we calculate $V_{eff}$ and $\rho$ for modes with radial orders varying from 1 up to 8. Fig.~\ref{fig:S2}a shows the results normalized by the values obtained for the first order mode. While $V_{eff}$ changes by at most 20\%, $\rho$ increases greatly, almost 16 times for the 8$^{\text{th}}$ order mode. Hence the \gnot dependence presented on Fig. 1e of the main text.
	
	The behavior of $\rho$ can be explained by the evanescent field of different radial order modes as a function of the distance from the disk border (Fig.~\ref{fig:S2}b). Although the field intensity for each order may vary at the edge of the disk, at the closest paddle's position (dashed line in Fig.~\ref{fig:S2}b) the decay rate difference results in higher field intensities for higher order modes.
	
	We note that the surface integral in $\rho$ should be calculated on all the surfaces of the two paddles but, because of the field decay rate, only the surface closest to the disk contributes significantly to \gnot. This is the reason why the measure of \gnot allows us to determine the better balanced device in our sample. That is because only for this device the amount of amplitude of motion of the front paddle is approximately the same for both S and AS modes.
	
	Nevertheless, one may ask why we measure our devices with the mode indicated in Fig.~\ref{fig:S1} if there are at least two other modes with higher radial order, hence higher \gnot. The answer is because the signal we obtain doesn't depend only on \gnot but also on the optical line-width, and we observe that the higher the mode order the larger the line-width becomes. Then, as we mentioned on the main text, we chose that particular mode for it was the one that yielded the best signal for our samples.
	
	\section{\label{sec.:S3}Coupled lumped oscillators model}
	
	\begin{figure}[h!tb]
		\center
		\includegraphics[scale=0.7]{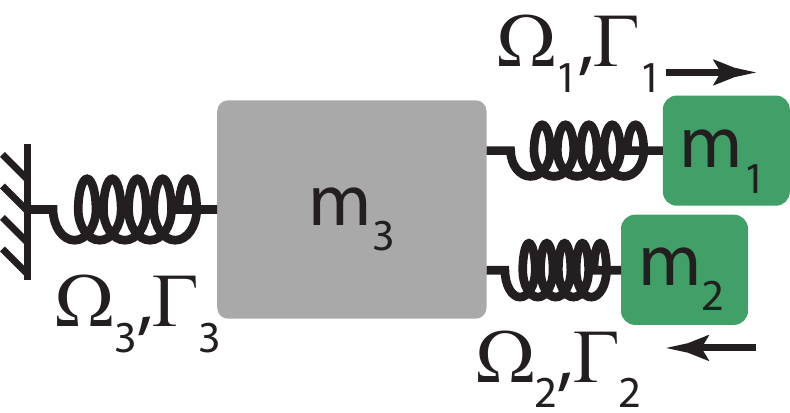}
		\caption{\label{fig:S4} Mass-spring lumped coupled oscillators model schematics.}
	\end{figure}
	
	We used a mass-spring lumped coupled oscillators model to explain the tuning fork effect that suppresses anchor loss in our devices. The model consists of two mass-spring oscillators (m$_1$ and m$_2$ in Fig.~\ref{fig:S4}), which represent the two paddles in our devices, coupled to a third mass-spring oscillator (m$_3$ in Fig.~\ref{fig:S4}), which represents the pedestal of our devices. We also include a direct coupling between oscillators 1 and 2, besides the indirect coupling by oscillator 3.
	
	We use the following simplified mathematical model for this system:
	\begin{eqnarray}
	\label{eq:S5}
	\frac{d x_1}{d t} &=  \mathfrak{i} \Omega_1 x_1 + \mathfrak{i} \frac{\kappa_{13}}{2} x_3 + \mathfrak{i} \frac{\kappa_{12}}{2} x_2 \nonumber\\
	\frac{d x_2}{d t} &=  \mathfrak{i} \Omega_2 x_2 + \mathfrak{i} \frac{\kappa_{23}}{2} x_3 + \mathfrak{i} \frac{\kappa_{21}}{2} x_2 \\
	\frac{d x_3}{d t} &=  \mathfrak{i} \Omega_3 x_3 + \mathfrak{i} \frac{\kappa_{31}}{2} x_1 + \mathfrak{i} \frac{\kappa_{32}}{2} x_2\nonumber
	\end{eqnarray}
	\noindent
	where we allow the frequencies $\Omega_{i} = \Omega_{i}+\mathfrak{i}\Gamma_{i}$ to be complex, $\kappa_{12}=\kappa_{21}$ is the coupling between resonators 1 and 2 and $\kappa_{13}=\kappa_{31}=\kappa_{23}=\kappa_{32}$ are the couplings of oscillators 1 and 2 with 3, $t$ is the time and $x_{i}$ are the amplitudes of motion of the oscillators.
	
	\begin{figure}[htb]
		\center
		\includegraphics[scale=1.0]{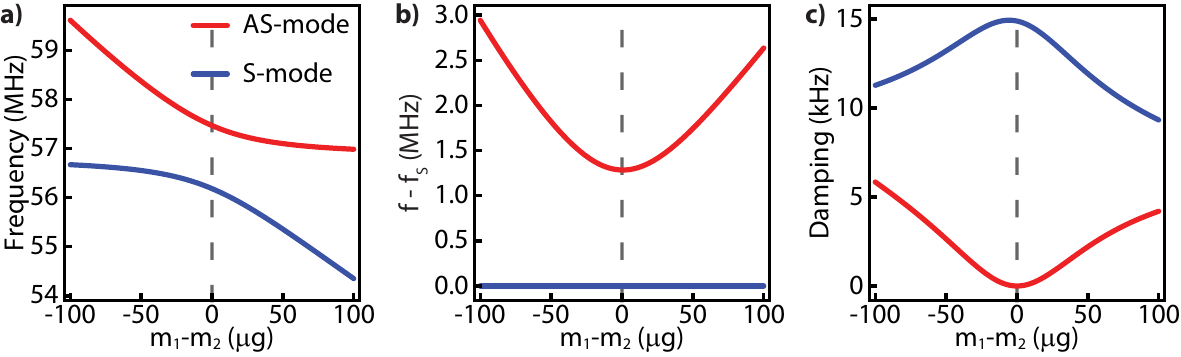}
		\caption{\label{fig:S5} \textbf{a)} Frequency of symmetric (S~\textemdash~blue) and anti-symmetric (AS~\textemdash~blue) modes obtained from the analytical model, varying the mass of oscillator 2. \textbf{b)} Same as \textbf{a} with frequency shifted with respect to the S mode resonance. \textbf{c)} Damping of S and AS modes obtained from the analytical model.}
	\end{figure}
	
	Solving the system of equations \ref{eq:S5} we obtain a set of three complex coupled mode eigenfrequencies, two for the symmetric (S) and anti-symmetric (AS) combinations of modes of the oscillators 1 and 2, and a third for the pedestal (not shown). By varying the mass of one of them (e.g. $m_2$) we are able to reproduce the expected avoided crossing of the coupled mode resonant frequencies and the shape of the damping rate observed in our FEM simulations (Fig.~\ref{fig:S5}). With the eigenvectors of this system we are able to reproduce the $\delta$ dependency of \gnot, as presented in figure 3d of the main text. Note that the example of solutions shown in Fig.~\ref{fig:S5} have minimum AS-mode damping when $m_{1} = m_{2}$, hence we add a factor to one of the resonances to take into account the symmetry braking caused by the pedestal in the real devices.
	
	The frequencies of oscillators 1 and 2 (paddles) are estimated as those of doubly clamped silicon beams' first flexural mode:
	\begin{equation}
	\Omega_{1,2} = \alpha_1^{2} \sqrt{\frac{Y t^{3} w}{12 m L^{3}}},
	\end{equation}
	\noindent
	where $\alpha_1 = 4.73$ is a constant derived from the beam theory~\cite{Cleland2003}, $Y$ is the Young's modulus of silicon, $t$ is the string width (200~nm), $w$ is the SOI thickness (220~nm) and $L$ the string length (4~\textmu m). We artificially modify the mass ($m$) to account for the extra mass the paddles add to the strings, without any changes to the elastic properties. We also included a scaling constant to this frequency to account for minor deviations from the experimental values, which does not affect the general behavior of coupled mode frequencies and damping rates. The frequency of oscillator 3 (pedestal) is estimated to be 76~MHz from FEM simulations of the device without the paddles and nanostrings.
	
	This model takes into account losses due to radiation to the substrate ($\Gamma_{3}$ and $\kappa_{i3}$) and intrinsic channels ($\Gamma_{1,2}$), but it does not consider the nature of the intrinsic losses neither does it account for their temperature dependence. We use this model to fit our experimental asymmetry ($\delta$ in the main text) dependent data for both room and low temperature, extracting from it the expected material and/or surface limited Q's for those conditions. Table \ref{tab:S1} resumes the resulting parameters that best fit our data, which were used to plot the curves in Fig. 3c,e,f of the main text.

	\begin{table}[h!tb]
		\center
		\caption{\label{tab:S1}Fitting parameters of a three coupled lumped mass-spring oscillators model}
		\begin{tabular}{|c|c|c||c|c|}
			\hline
			&\multicolumn{2}{|c||}{\textbf{Room temperature}} & \multicolumn{2}{|c|}{\textbf{Low temperature}}\\
			\hline
			Parameter & Mean & Std. Err. & Mean & Std. Err.\\
			\hline
			$\kappa_{12}$ & 0.647 MHz & 18.7~\% & 0.647 MHz & 18.7~\%\\
			\hline
			$\kappa_{13}$ & 3.99 MHZ & 18.7~\% & 3.99 MHZ & 18.7~\%  \\
			\hline
			Q$_{3}$ & 323 & 5.27~\% & 320 & 5.5~\%  \\
			\hline
			Q$_{1,2}$ & 7376 & 1.67~\% & 27867 & 4.6~\%  \\
			\hline
		\end{tabular}
	\end{table}
	
	\section{\label{sec.:S4} Thermoelastic damping~\textemdash~TED}
	
\begin{figure}[h!tb]
		\center
		\includegraphics[scale=1]{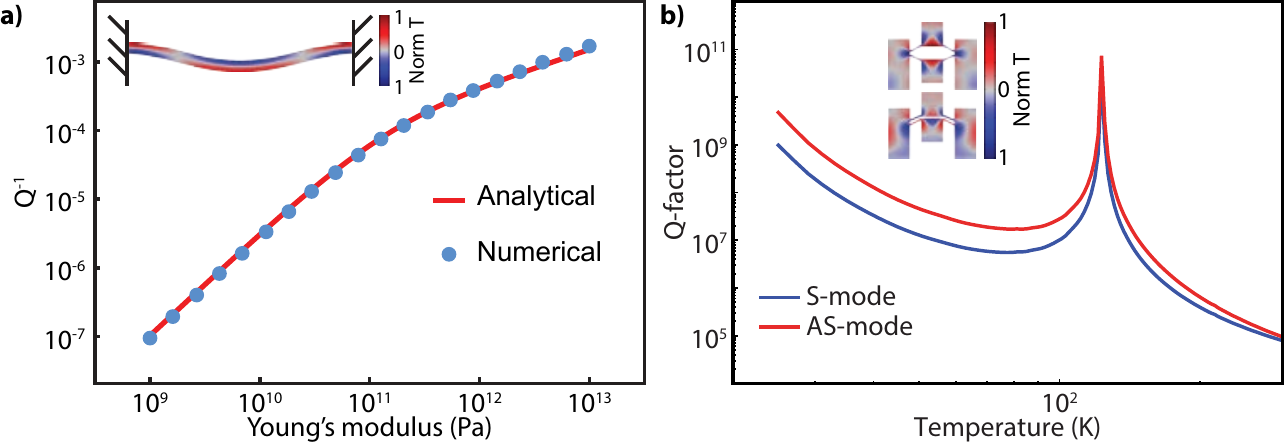}
		\caption{\label{fig:S6} a) Comparison of numeric and analytical models to TED on a doubly clamped beam; inset: side-view of numerical thermal distribution. b) Numerical solution of temperature dependent TED limited Q for the S and AS modes of the paddles.}
	\end{figure}

	To assess the role of thermoelastic damping (TED) in our devices we performed FEM simulations of this effect. We verified the validity of our FEM model by comparing the numerical solution of a doubly clamped beam to the analytical model presented by Roukes and Lifshitz~\cite{Lifshitz2000}. The result is presented in Fig.~\ref{fig:S6}a, where we artificially varied the material Young's modulus in both models.
	
	Once the FEM model was validated we applied it to our devices, using material properties for temperatures varying from 300~K down to 25~K (see section \ref{sec.:S6}). The results for TED lmited Q's are in Fig.~\ref{fig:S6}b, where we observe that at room temperature the minimum Q is 80k and it increases to almost $10^{10}$ at 25~K. Anyhow, this effect's model predicts Q limits much higher than the typical measured in our devices, at all temperatures.
	
	\section{\label{sec.:S6}Temperature dependent Silicon properties and the Akhiezer model}
	
\begin{figure}[htb]
		\center
		\includegraphics[scale=1]{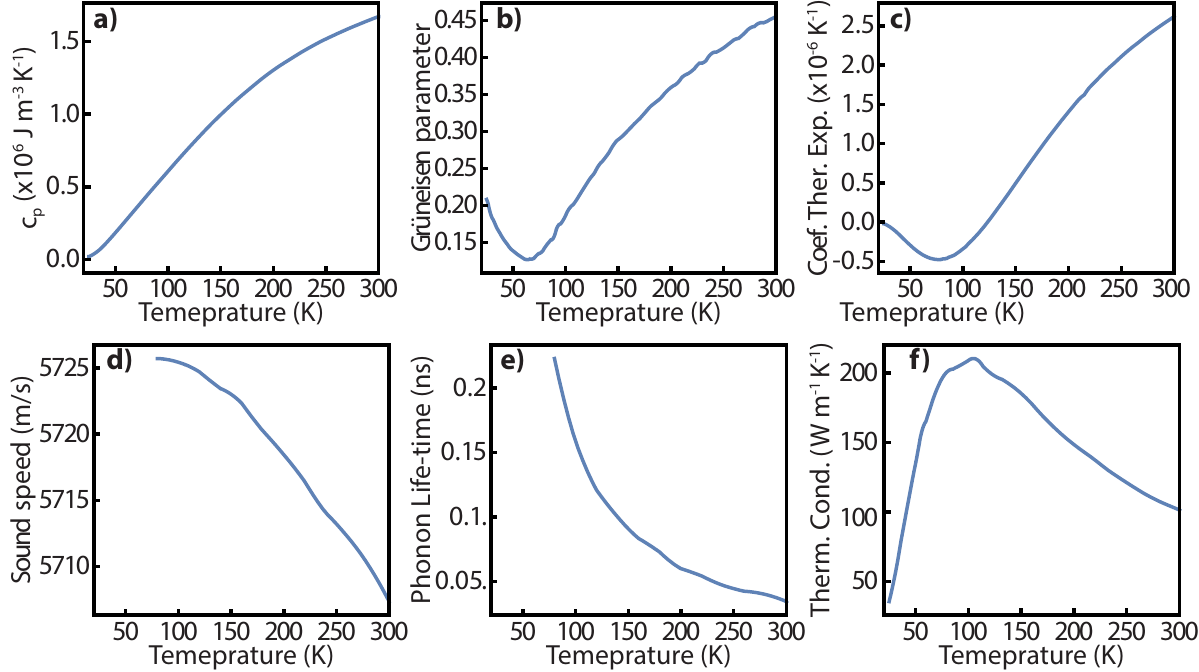}
		\caption{\label{fig:S9}\textbf{Silicon properties versus temperature.} \textbf{a)} Heat capacity~\cite{Desai1986}; \textbf{b)} Gr\"uneisen parameter~\cite{Philip1983}; \textbf{c)} Coeffcient of thermal expansion~\cite{Lyon1977}; \textbf{d)} Average Debye sound speed~\cite{Lambade1995}; \textbf{e)} Mean thermal phonon life-time~\cite{Lambade1995}; \textbf{f)} Thermal conductivity~\cite{Asheghi1998}.}
	\end{figure}

	Fig.~\ref{fig:S9} summarizes the properties of Silicon used to estimate TED and AKE mechanical Q limits as a function of temperature. For TED we use SOI thermal conductivity from Asheghi et al~\cite{Asheghi1998}, heat capacity from Desai~\cite{Desai1986} and coefficient of thermal expansion from Lyon et al~\cite{Lyon1977}. For AKE we use the heat capacity from Desai~\cite{Desai1986}, sound speed and thermal phonon life-time from Lambade et al~\cite{Lambade1995} and the Gr\"uneisen parameter from Philip and Breazeale~\cite{Philip1983}.
	
	For the AKE damping we used the expression derived by Woodruff and Ehrenreich~\cite{Woodruff1961a}.
	\begin{equation}
	\Gamma_{\text{AKE}}= \frac{\gamma^2 c_p T}{3 \rho v_{D}^{2}}\ 4 \pi^{2} f_m^2 \tau
	\end{equation}
	\noindent
	where for sound speed we use the Debye average of the values available for phonons propagating in the $[110]$ direction ($v_{D}$). For thermal phonon life-time we use the values for those interacting with acoustic phonons propagating in the $[110]$ direction with polarization in the $[1\bar{1}0]$ direction. We do not use the conductivity due to the problems in its relation to the phonon life-time discussed by Ilisavskii~\cite{Ilisavskii1985}. We note that these values are not available for temperatures below 80~K, hence we extrapolate these parameters to obtain estimates below this temperature.
	
	\section*{Funding}
	
	This work was supported with funds from FAPESP (grants: 2008/57857-2, 2012/17610-3, 2012/17765-7) and CNPq (grant 153044/2013-6).

\input{loss_channels_arXiv.bbl}	

\end{document}

%% file: loss_channels_arXiv.bbl
%